\documentstyle[aps,multicol,prl,epsfig]{revtex}

\begin{document}
\begin{title}
{\bf 
Feedback Loops Between Fields and Underlying Space Curvature: an Augmented
Lagrangian Approach}
\end{title}

\author{P.G. Kevrekidis$^1$, F.L. Williams$^1$, A.R. Bishop$^2$, 
I.G. Kevrekidis$^3$, and B.A. Malomed$^4$}
\address{$^1$ Department of Mathematics and Statistics, 
University of Massachusetts, Amherst, MA 01003-4515, USA \\ 
$^2$ 
Theoretical Division, MS B210,
Los Alamos National Laboratory, Los Alamos, NM 87545, USA \\
$^3$ Department of Chemical Engineering, Princeton University,
6 Olden Str. Princeton, NJ 08544 \\
$^4$ Department of Interdisciplinary Studies, Faculty of Engineering, Tel
Aviv University, Tel Aviv 69978, Israel \\
}
\maketitle

\begin{abstract}
We demonstrate a systematic implementation of coupling between
a scalar field and the geometry of the space
(curve, surface, etc.) which carries the field. This naturally
gives rise to a feedback mechanism
between the field and the geometry. We develop a systematic model for the
feedback in a general form, inspired by a specific implementation in the
context of molecular dynamics (the so-called Rahman-Parrinello molecular
dynamics, or RP-MD). We use a generalized Lagrangian that allows for the
coupling of the space's metric tensor (the first fundamental form) to the
scalar field, and add terms motivated by RP-MD. We present two
implementations of the scheme: one in which the metric is only
time-dependent [which gives rise to ordinary differential equation (ODE) for
its temporal evolution], and one with spatio-temporal dependence [wherein
the metric's evolution is governed by a partial differential equation (PDE)].
Numerical results are reported for the (1+1)-dimensional model with a
nonlinearity of the sine-Gordon type.
\end{abstract}


\vspace{5mm}


\begin{multicols}{2}

Recently, much attention has been focused on soft-condensed-matter objects,
such as vesicles, microtubules, and membranes \cite{av1,av2,gaid1,gaid2}.
Many nanoscale physical systems, including nanotubes and electronic and
photonic waveguide structures \cite{gaid3,gaid4}, have nontrivial geometry
and are influenced by substrate effects. These classes of systems, many of
which are inherently nonlinear, raise the question of the interplay between
nonlinearity and a substrate with variable curvature. Of particular interest
is a possibility of developing curvature in the substrate due to forces
generated by the nonlinear field. The resulting curvature can in turn affect
the field.

There is an increasing body of literature dealing with the interplay of
nonlinearity and a curved substrate. Usually, however, the substrate
geometry is assumed to be {\it fixed}, see, e.g.,  \cite{curv}.
Nevertheless, for many applications, ranging from condensed matter to optics
to biophysics, it is relevant to introduce  models that admit
a flexible substrate, which is affected by the field(s) that it carries, as
well as feeding back into the field dynamics. In this situation, equations for
the fields in a nonlinear system abutting on the flexible substrate should
include both the field dynamics proper and the feedback coupling to the
substrate. Equations for the evolution of the substrate should in turn be
affected by the evolution of the field. A prototypical physical example of
this type is Euler buckling \cite{antm}, where the evolution of a thermal
profile causes the underlying surface to buckle (and hence locally modify
its curvature).

In a discrete setting, a model of this type has recently been presented in
 \cite{kmb}. However, it was limited to a system of masses coupled by
nonlinear springs. Some studies have also been performed in a special case
of the continuum limit of classical spin systems (such as the Heisenberg
chain) coupled to the curvature; geometric frustration was found to arise in
such settings \cite{dand}.

About twenty years ago, a problem similar to the theme of our study was
examined in the context of molecular dynamics studies of
structural transitions in crystals. 
In particular, in a series of papers \cite{RP}, Rahman and
Parrinello introduced a new idea for studying such transitions by means of
an augmented Lagrangian that would account for the degrees of freedom of the
``box'' (the cell) in which the MD particles lie. In
studying the time evolution of the box dynamics (naturally obtained through
the Euler-Lagrange equations of the augmented Lagrangian for the box degrees
of freedom), they were able to identify structural transitions (under
external shear) from square to hexagonal patterns, fcc to bcc etc. We will
hereafter refer to this technology as the RP-MD method. The relevant
Lagrangian for the particles and the box in this case reads: 
\begin{equation}
L=\frac{1}{2}\sum_{i}m_{i}{\bf \dot{s}}_{i}G{\bf \dot{s}}_{i}-
\sum_{i,j>i}V(r_{ij})+\frac{1}{2}W{\rm Tr}(\dot{f}^{{\rm T}}\dot{f}),
\label{rpeq1}
\end{equation}
where $m_{i}$ is the mass of the $i$-th particle, ${\bf \dot{s}}_{i}$ is its
vectorial velocity, the spatial part $G$ of the spatiotemporal metric tensor
may be represented in terms of another matrix $f$ as $G=f^{{\rm T}}f$ ($G$
is positive definite). $^{{\rm T}}$ and ${\rm Tr}$ denote the transposition
and trace respectively, 
$r_{ij}$ and $V$ are distances between the particles and the
potential of interaction between them, and $W$ is an effective mass of the
box.

Our purpose in this work is to extend the RP-MD methodology
to the case of a continuum scalar field, 
coupled to either a spatially averaged geometric
characteristic (``average curvature''), which will give rise to an ODE, or
to a spatiotemporal curvature field, that will generate a PDE. 
The continuum field
may represent, e.g., a chemical concentration propagating over a membrane, 
or a salt solution, causing the swelling of a polymer gel \cite{yy}, 
or an envelope
wave of the electric
field in nanosystems.
The
spatiotemporal metric is assumed to have the simple form, 
\begin{equation}
g=\left( 
\begin{array}{cc}
-1 & 0 \\ 
0 & G
\end{array}
\right) .  \label{rpeq2}
\end{equation}
We will first consider the general case, where $G$ is a $d\times d$ matrix, 
$d$ being the space dimension.

One can define a field-type generalization of the RP-MD model, with a scalar
field $\phi $, as follows: 
\begin{equation}
L=\int d^{d}x\left[ -g^{ij}\frac{\partial \phi }{\partial x^{i}}\frac{
\partial \phi }{\partial x^{j}}-V(\phi )\right] +\frac{1}{2}W{\rm Tr}(\dot{f}
^{{\rm T}}\dot{f}),  \label{rpeq3}
\end{equation}
where $V(\phi )$ is a potential governing the nonlinear evolution of the field 
$\phi $ and $f=f(t)$ (only).
If $f$ (and hence $G$) is a function of both spatial coordinates and
time, an elastic-energy term \cite{elastic} should be added to the Lagrangian 
(\ref{rpeq3}), so that it becomes 
\[
L=\int d^{d}x\left[ -g^{ij}\frac{\partial \phi }{\partial x^{i}}\frac{
\partial \phi }{\partial x^{j}}-V(\phi )\right] + 
\]
\begin{equation}
\int d^{d}x\left\{ \frac{1}{2}W\left[ {\rm Tr}(\dot{f}^{{\rm T}}\dot{f})-
\frac{1}{2}{\rm Tr}(\frac{\partial f^{{\rm T}}}{\partial x_{i}}\frac{
\partial f}{\partial x^{i}})\right] \right\} .  \label{rpeq4}
\end{equation}


An objective of this brief report is to propose two models that include the
feedback to the curvature, and, simultaneously,  admit as particular
solutions the (unperturbed) solutions for the flat 
(original) metric. In particular, we choose $G=1+f^2$ 
(so that the metric is positive definite and for $f=0$ has as a special
case the original Minkowskian metric). Notice a slight deviation in
our choice from the RP case of $G=f^2$ \cite{choice}. Although the
formulation of Eqs. (\ref{rpeq3})-(\ref{rpeq4}) is very general, we
hereafter focus on the $d=1$ case that we will examine
in more detail.

Assuming initially that $f=f(t)$ only 
(e.g., including only an effect of the ``mean
curvature'' on the scalar-field dynamics), the Lagrangian (\ref{rpeq3})
becomes 
\begin{equation}
L=\frac{1}{2}W\dot{f}^{2}+\int dx\left[ \frac{1}{2}\left( \frac{\partial
\phi }{\partial t}\right) ^{2}-\frac{1 + f^{2}}{2}\left( \frac{\partial \phi }
{\partial x}\right) ^{2}-V(\phi )\right] .  \label{rpeq6}
\end{equation}
Then, the resulting equations of motion (to which we will hereafter refer
as model A) are
\begin{eqnarray}
\phi _{tt} &=&(1+f^{2})\phi _{xx}-\frac{\partial V}{\partial \phi }
\label{rpeq13} \\
Wf_{tt} &=&-f\int \phi _{x}^{2}dx,  \label{rpeq14}
\end{eqnarray}
where the subscripts stand for the corresponding partial derivatives.

Notice that the function $f$ is directly related to the
scalar curvature of the 1-d space. In particular, the Ricci scalar, 
which is $R=2R_{1212}/det(g)$ \cite{weinberg} 
in the general case, in the 1-d case
is $R=-2 \tilde{f}_{tt}/\tilde{f}$, where $\tilde{f}=\sqrt{1+f^2}$.

On the other hand, for a metric with both spatial and temporal
dependence (e.g., for $f=f(x,t)$), 
one arrives at the following Lagrangian: 
\[
L=\int dx\left[ \frac{1}{2}\left( \frac{\partial \phi }{\partial t}\right)
^{2}-\frac{1 + f^{2}}{2}\left( \frac{\partial \phi }{\partial x}\right)
^{2}-V(\phi )\right] + 
\]
\begin{equation}
\int dx\left[ \frac{W}{2}\left( \frac{\partial f}{\partial t}\right) ^{2}-
\frac{W}{2}\left( \frac{\partial f}{\partial x}\right) ^{2}\right] . 
\label{add1}
\end{equation}
The ensuing coupled equations for the scalar field and the curvature 
(to which we will refer as model B) are 
\begin{eqnarray}
\phi _{tt} &=&\left( (1+f^{2})\phi _{x}\right) _{x}-\frac{\partial V}
{\partial \phi }  \label{rpeq15} \\
Wf_{tt} &=&Wf_{xx}-f\phi _{x}^{2}.  \label{rpeq16}
\end{eqnarray}


As a particular application of models A and B, 
we examine the physically ubiquitous
sine-Gordon (sG) potential, $V(\phi )=1-\cos (\phi )$ \cite{eilbeck}. It is
clear that Eqs. (\ref{rpeq13})-(\ref{rpeq14}) and
(\ref{rpeq15})-(\ref{rpeq16}) have particular
solutions with $f=0$, for which the latter equation of each pair is satisfied
trivially, while the former reduces exactly to the sG equation. Basic
solitary-wave solutions of the sG equation are the topological soliton
(kink), 
\begin{equation}
\phi _{{\rm k}}(x,t)=4\tan ^{-1}\left[ \exp \left( \gamma
(x-x_{0}-vt)\right) \right] ,  \label{kink}
\end{equation}
where $v$ is its velocity, $\gamma =(1-v^{2})^{-1/2}$ is the Lorentz factor,
and $x_{0}$ is the initial position of the kink's center, and the breather, 
$\phi _{{\rm br}}(x,t)=4\tan^{-1} \{ \sqrt{(1-\omega ^{2})/\omega ^{2}}
\sin \left[ \omega \gamma (t-v(x-x_{0}))\right]  
\times \\
{\rm sech}\left[
\gamma \sqrt{1-\omega ^{2}}(x-x_{0}-vt)\right] \} $, where $\omega $
is the frequency of its internal oscillations ($0<\omega <1$).

\begin{figure}[tbp]
\centering
{\epsfig{file=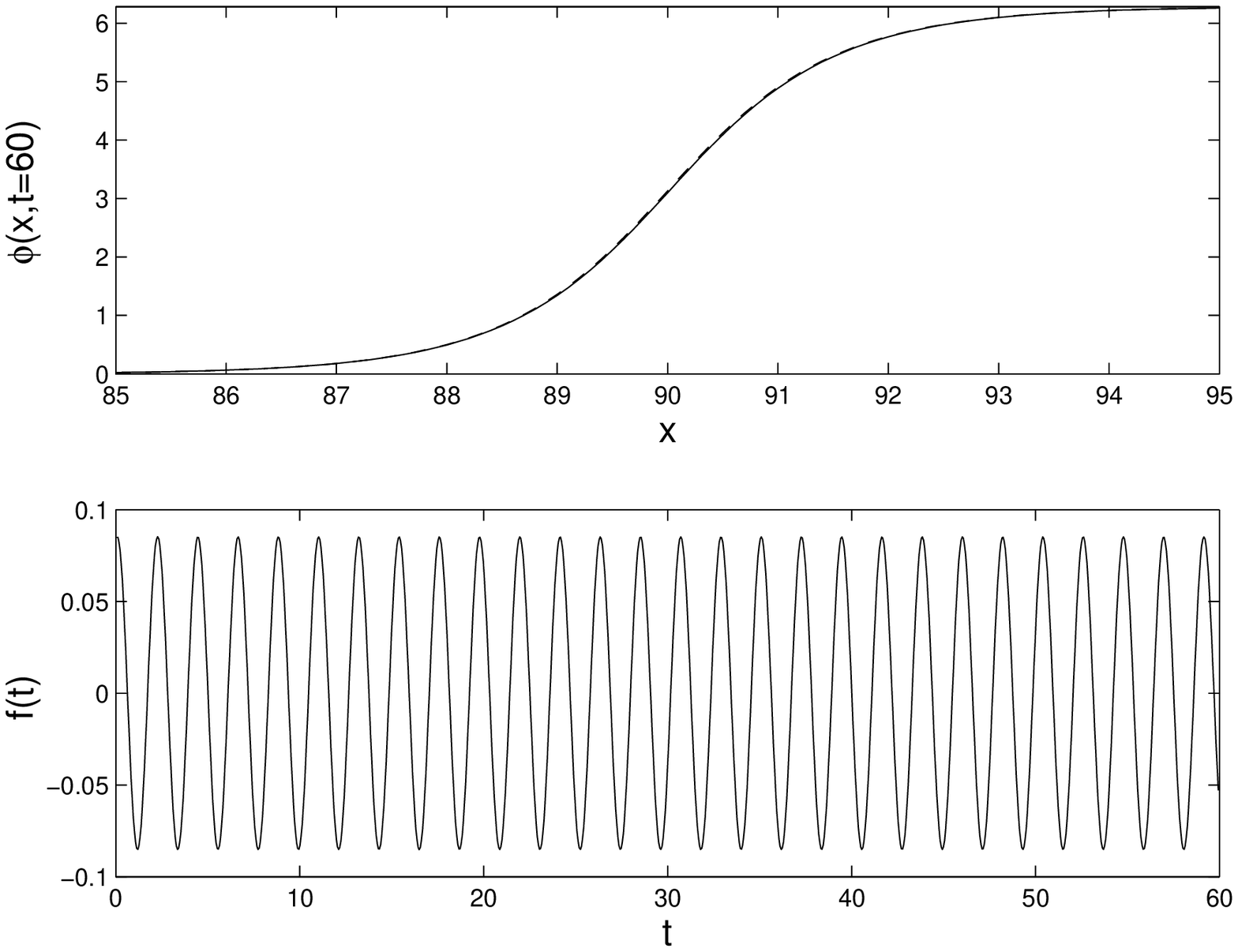, width=6.4cm,angle=0, clip=}}
\centering
{\epsfig{file=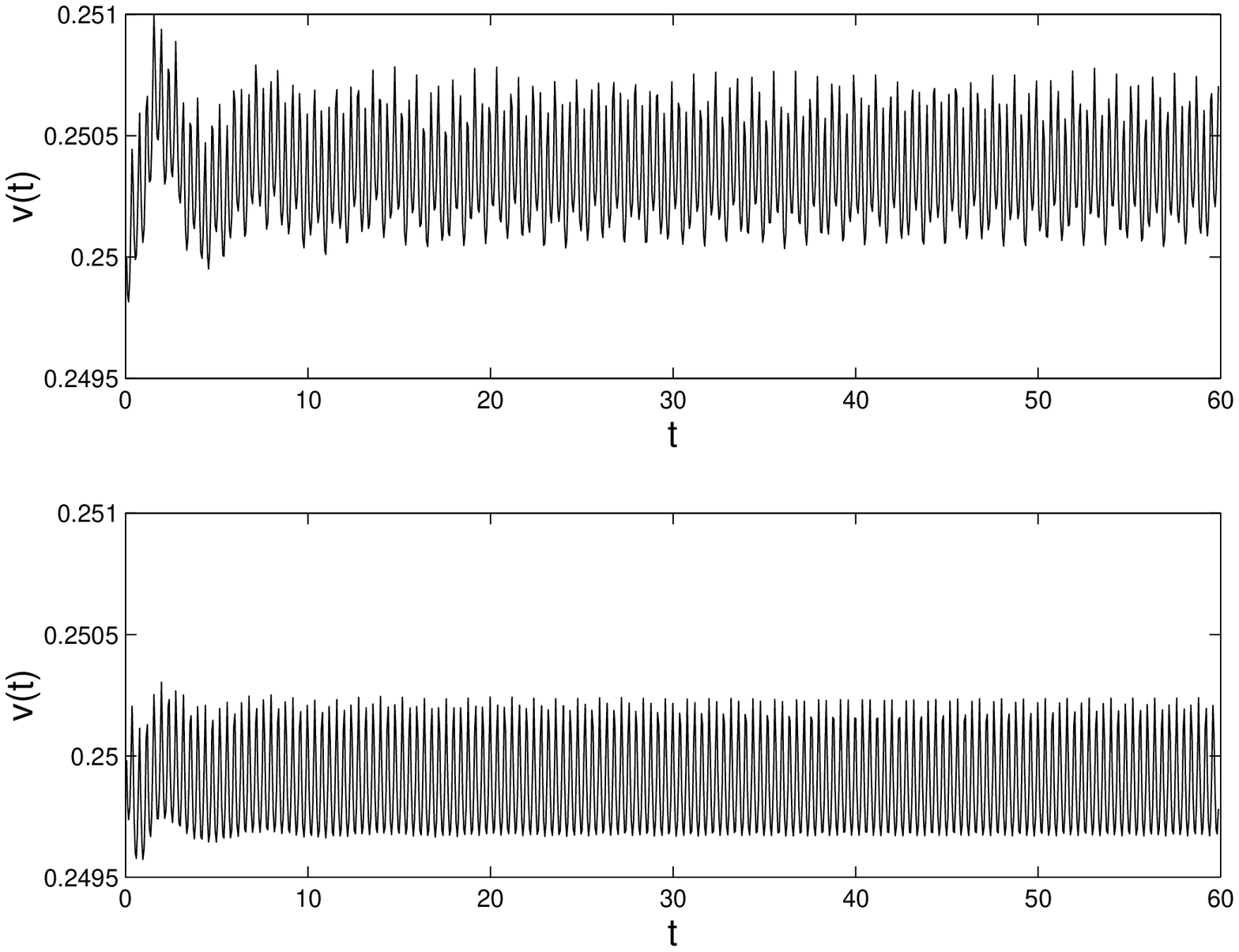, width=6.4cm,angle=0, clip=}}
\caption{The top left panel shows the kink's spatial profile at $t=60$ in
model A. The initial kink profile (centered around the new center
position) is shown by the dashed line, and is practically indistinguishable
from the solution at $t=60$, indicating that the kink maintains its shape.
The bottom left panel shows smooth oscillations of $f(t)$. The top and
bottom right panels show, respectively, the kink's velocity $v(t)$ vs. $t$,
and the same quantity but for $f=0$.}
\label{rfig2}
\end{figure}

The results of the interaction of the kink with the curvature in model A
are shown in Fig. \ref{rfig2} \cite{ww}. 
The curvature variable $f$, initialized with
a small random value, performs smooth oscillations with a frequency of 
$\omega \approx 2.866$. Notice that this is natural in this case, since the
kink has an approximately fixed ``mass'', $M_{k}=\int_{-\infty }^{+\infty
}u_{x}^{2}dx$, which can be found to be $M_{k}=8.247$ for the velocity $%
v\approx 0.25$). Then, Eq. (\ref{rpeq14}) predicts the frequency of these
oscillations $\omega \approx \sqrt{M_{k}}=2.871$, which is very close to the
above-mentioned numerically exact value. Fast small-amplitude oscillations
of the kink's velocity, observed in Fig. \ref{rfig2} both with and without
the curvature, are due to ``hopping'' over sites of a lattice (with spacing 
$h=0.1$) employed in the numerical scheme which solves Eq. (\ref{rpeq14}).
Notice, however, that in the top panel the mean velocity is $\approx 0.2503$,
while in the bottom panel it is $\approx 0.2499$, hence the curvature
oscillations increase the kink's velocity. This may be anticipated due to
the presence of the positive definite factor $1+f^{2}$ in front of $\phi
_{xx}$ in Eq. (\ref{rpeq13}), which is expected to renormalize $v^{2}$.

\begin{figure}[tbp]
\centering
{\epsfig{file=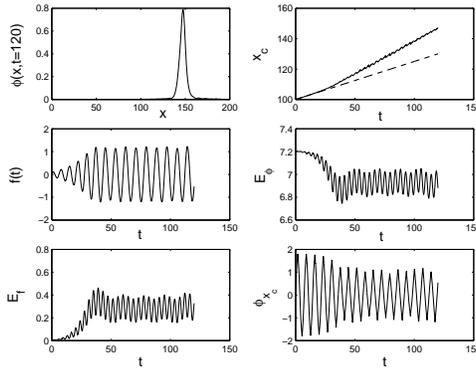, width=6.4cm,angle=0, clip=}}
\caption{The top left and right panels show, respectively, the breather in
model A at the end of the simulation period, $t=120$, and its position
vs. time (solid), as compared to that which it would have moving at the
initial velocity, $v=0.25$ (dashed). The middle left panel shows the time
evolution of $f(t)$, while the middle right and bottom left panels present
exchange between the curvature-mode's energy, $E_f= f_t^2/2 + (f^2/4) \int 
\protect\phi_x^2 dx$, and the rest of the energy, $E_{\protect\phi}=E-E_f$.
Finally, the bottom right panel shows the time evolution of the field 
$\protect\phi$ at the center of the breather.}
\label{rfig3}
\end{figure}

The curvature-breather interaction in model A is shown in Fig. \ref
{rfig3}. The frequency of the breather does not change significantly (it
fluctuates between $0.87$ and $0.93$), but its amplitude decreases
substantially (by more than $50\%$), resulting in its becoming much more
mobile (the velocity increases to $\approx 0.425$ from the initial value 
$0.25$).

\begin{figure}[tbp]
\centering
{\epsfig{file=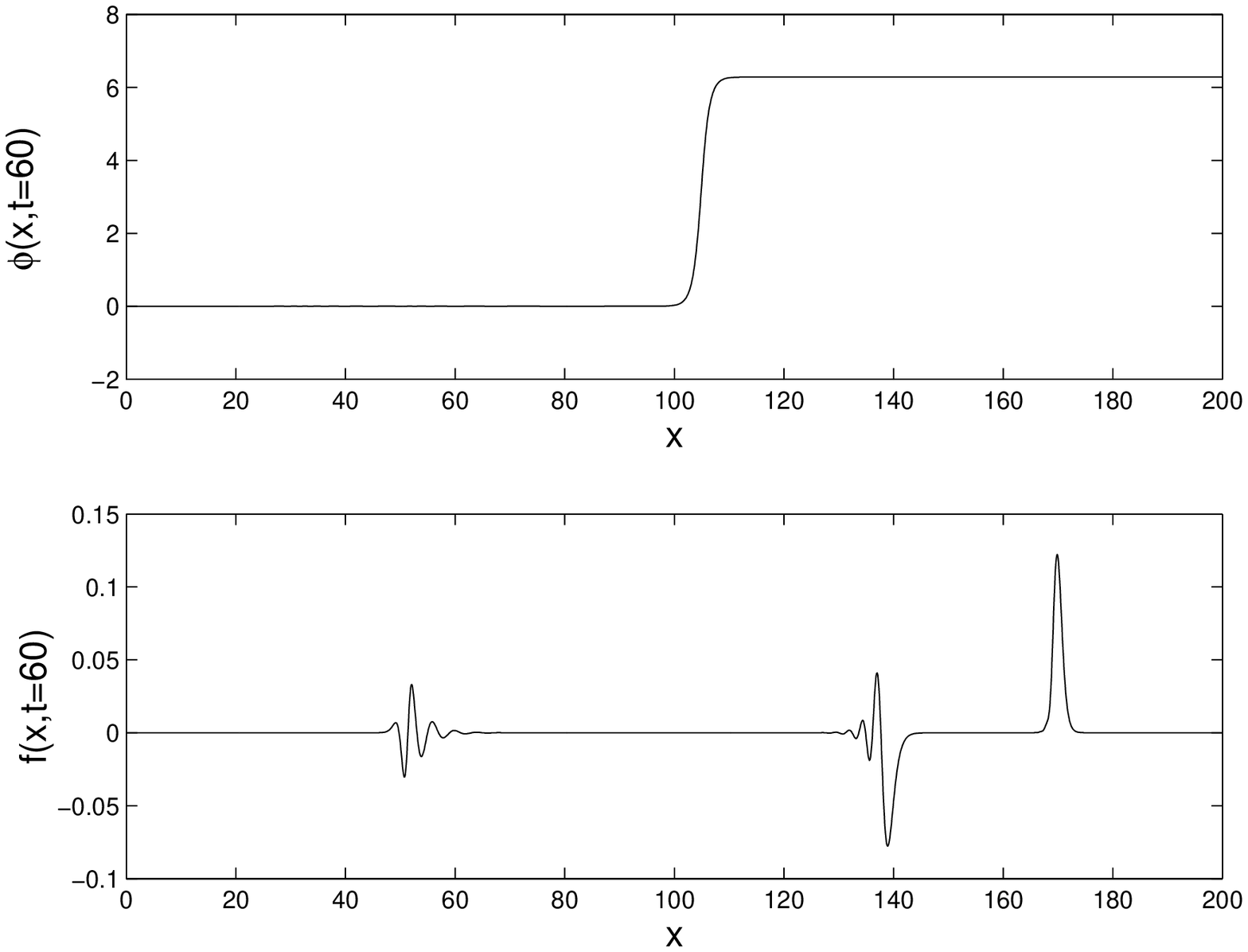, width=6.4cm,angle=0, clip=}}
\centering
{\epsfig{file=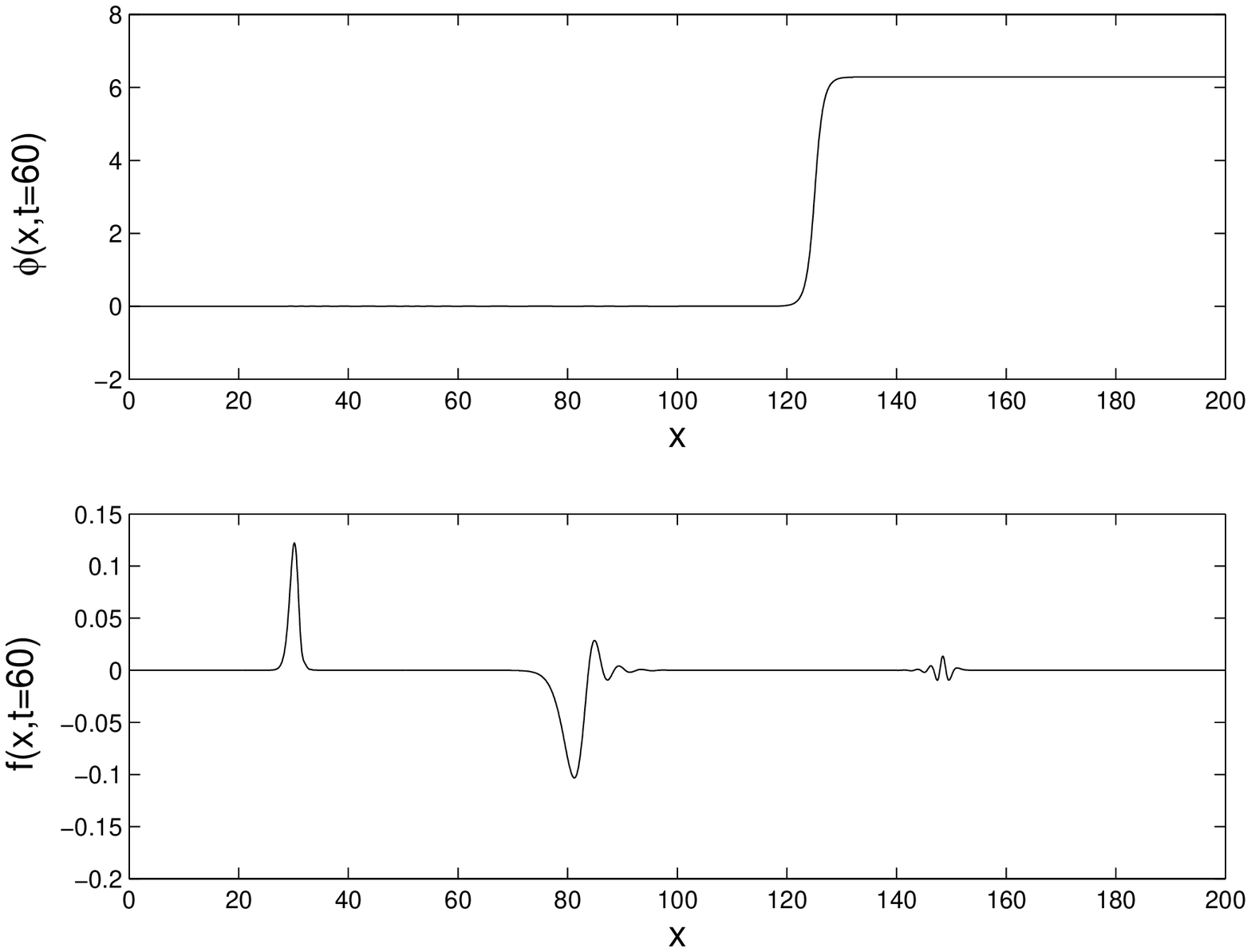, width=6.4cm,angle=0, clip=}}
\centering
{\epsfig{file=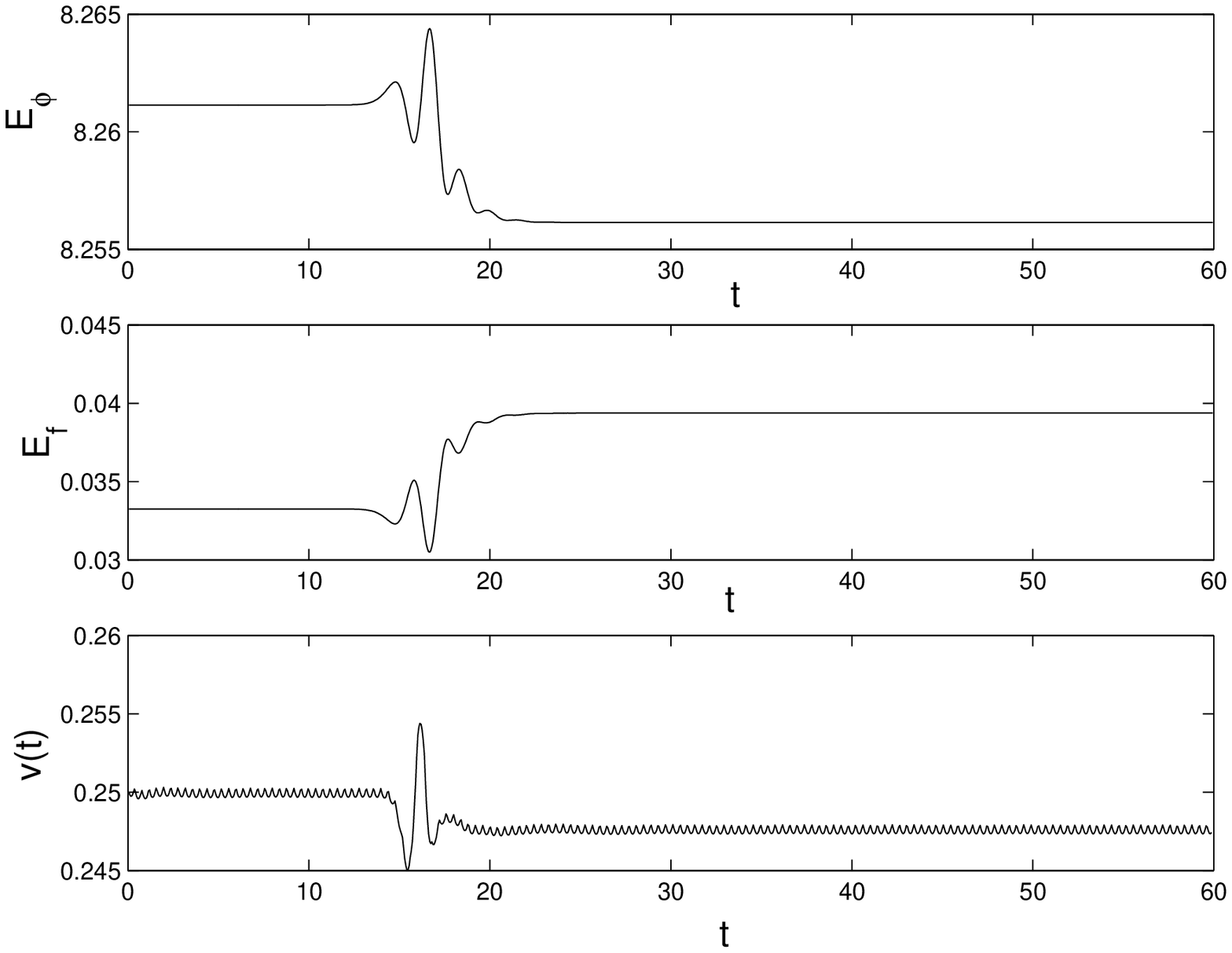, width=6.4cm,angle=0, clip=}}
\centering
{\epsfig{file=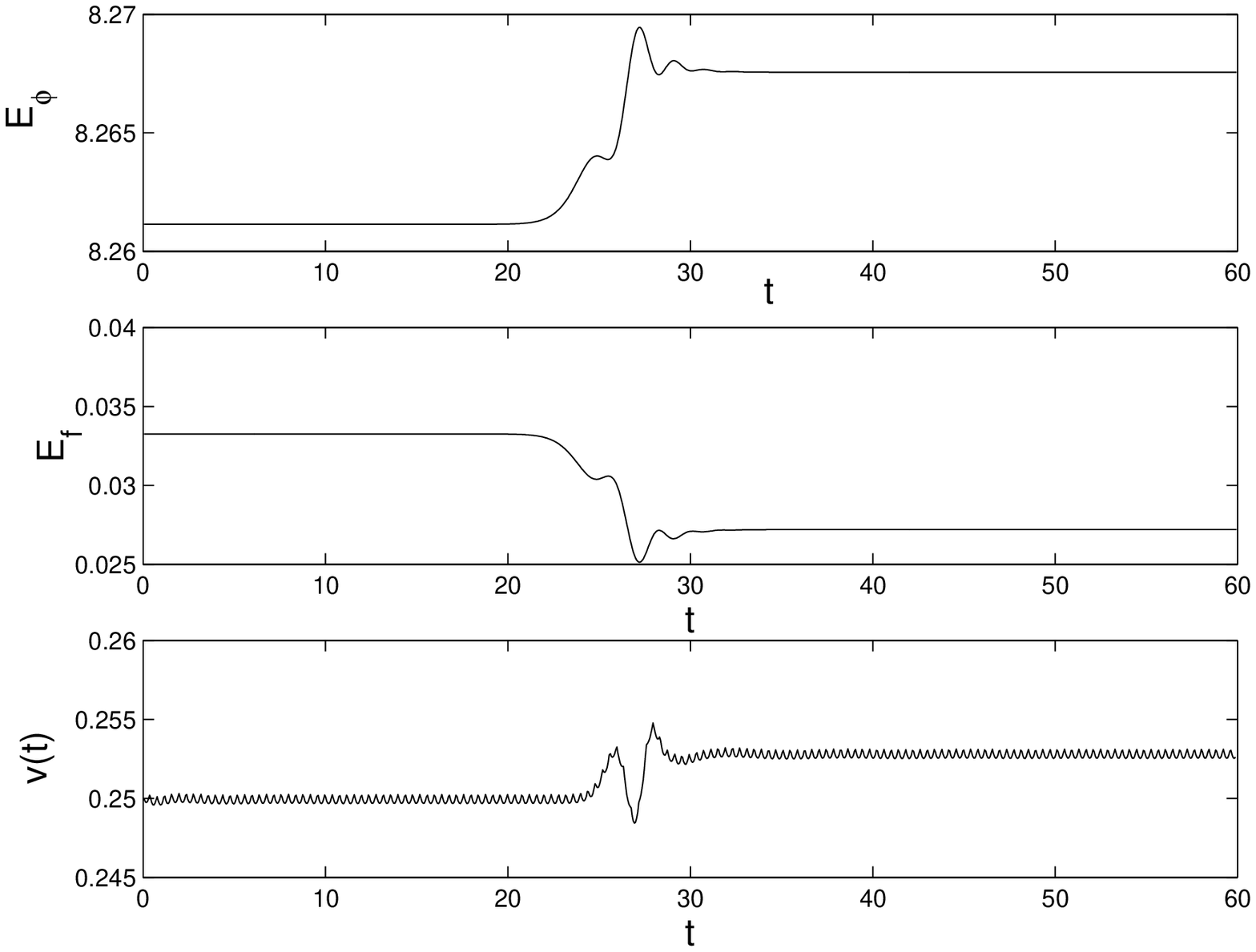, width=6.4cm,angle=0, clip=}}
\caption{The left part of the figure shows spatial profiles of the fields 
$\protect\phi$ and $f$ at $t=60$ in model B. It also shows the time
evolution of the energies $E_{\protect\phi}=E-E_f$ and $E_f=\int (1/2)
(f_t^2+f_x^2+ (f \protect\phi_x)^2/2 ) dx$, as well as the kink's velocity
as a function of time. The kink's center is initially at $x=90$, while the
center of the curvature pulse is at $x=110$. The right subplots show
counterparts of those in the left part, but for the case of the kink and
curvature pulse initially centered at $x=110$ and $x=90$, respectively.}
\label{rfig4}
\end{figure}

In model B, we examine collision of a kink with a localized pulse of the
substrate field $f$. For the breather, we have obtained results which are
qualitatively similar to those presented below for the kink. We create the
curvature pulse to the left or to the right of the kink. As $\phi _{x}$
vanishes far from the kink, the equation (\ref{rpeq16}) for $f$ becomes a
linear wave equation. Hence, we observe splitting of the pulse into left-
and one right-traveling ones. In the case where the kink is initially to the
left of the pulse, it collides with the left-propagating fragment of
the (split) pulse. In the opposite case,
the kink is eventually caught by the co-propagating right fragment of
the (split) pulse. Numerical
simulations shown in Fig. \ref{rfig4} demonstrate that the collision with
the counter-propagating pulse reduces the kink's velocity, while the
interaction with the co-propagating pulse gives rise to an increase of the
velocity. In particular, in the former case, the mean speed of the kink
after the collision is $\approx 0.2476$, while in the latter one, it
increases to $\approx 0.2527$. Notice that in both cases a small fragment of
the pulse that collides with the kink passes through it, while a larger
fraction of the pulse is reflected by it. The latter feature may be
explained by a momentum-balance analysis.

In conclusion, we have extended field theory in the spirit of
the Rahman-Parrinello
Molecular Dynamics technique. The resulting
equations couple the spatio-temporal evolution of the field to that of the
underlying curvature of the space which carries the field. Coupled equations
for the temporal or spatiotemporal evolution of the metric are obtained in a
general setting, and, as an example,
are solved together with the sine-Gordon field
equation. The purely temporal evolution of the metric has been found to
increase the velocity of the field solitons, while the model allowing
spatio-temporal evolution of the metric can induce both increase and
decrease of the velocity.

It would be particularly interesting to extend the models A and B to higher
dimensions. It is also worth studying how the local evolution of the curvature
affects kinematics and dynamics of the solitons, and to correlate such
observations with the behavior of reactant chemical concentrations in
chemical or biological environments with non-trivial geometry \cite{yannis}.

{\it Acknowledgements}: The authors acknowledge D. Maroudas for 
stimulating discussions. The support of NSF and AFOSR (IGK) and
NSF (DMS-0204585), UMass and the Clay Institute (PGK) is gratefully
acknowledged. Work at Los Alamos is supported by the US DoE.

\end{multicols}



\begin{references}
\bibitem{av1}  E. Sackmann, in R. Lipowsky and E. Sackmann (Eds.), The
Structure and Dynamics of Membranes: Handbook of Biological Physics, vol. 1
(Elsevier, Amsterdam, 1995).

\bibitem{av2}  U. Seifert, \newblock Adv. in Phys. {\bf 46}, 13 (1997).

\bibitem{gaid1}  W. Saenger. 
\newblock{\it Principles of Nucleic Acid
Structure} (Springer-Verlag: Berlin, 1984).

\bibitem{gaid2}  J. Meunier, D.
Langevin, and D. Boccara (Eds.),
{\it Physics of Amphiphilic Layers},  Springer-Verlag (Berlin, 1997).

\bibitem{gaid3}  Y. Imry. {\it Introduction to Mesoscopic Physics} (Oxford
University Press: New York, 1997).

\bibitem{gaid4}  C.M. Soukoulis
(Ed.), {\it Photonic Band Gaps and Localization}, Plenum Press 
(New York, 1993).

\bibitem{curv}  See e.g., Yu. B. Gaididei, S. F. Mingaleev, and P. L.
Christiansen, Phys. Rev. E {\bf 62}, R53 (2000); M. Ibanes, J. M. Sancho,
and G. P. Tsironis, Phys. Rev. E {\bf 65}, 041902 (2002);

\bibitem{antm} S.P. Timoshenko and S. Woinowsky-Krieger,
{\it Theory of plates and shells}, McGraw-Hill (New York, 1959)
S.S. Antman,
\newblock {\it Nonlinear problems of elasticity}, Springer-Verlag 
(New York 1995).

\bibitem{kmb}  P.G. Kevrekidis, B.A. Malomed and A.R. Bishop, \newblock
Phys. Rev. E {\bf 66}, 046621 (2002).

\bibitem{dand}  S. Villain-Guillot, R. Dandoloff, A. Saxena and A.R. Bishop,
\newblock Phys. Rev. B {\bf 52}, 6712 (1995); R. Dandoloff, S.
Villain-Guillot, A. Saxena and A.R. Bishop, \newblock Phys. Rev. Lett. {\bf
74}, 813 (1995).


\bibitem{RP}  see e.g., M. Parrinello and A. Rahman, \newblock Phys. Rev.
Lett. {\bf 45}, 1196 (1980). M. Parrinello and A. Rahman, \newblock J. Appl.
Phys. {\bf 52}, 7182 (1981).

\bibitem{yy} E. C. Achilleos, R. K. Prud'homme, I. G. Kevrekidis, K. N. Christodoulou and
K.R. Gee,
{ AIChE Journal}, {\bf 46}, 2128 (2000).

\bibitem{elastic} Elastic dynamics is a natural first approximation,
even though it is possible for the substrate  to have more 
complicated dynamical evolution laws (which can also be incorporated
in this setting).

\bibitem{choice} We also ran detailed simulations of the case with 
$G=f^2$. Here, the non-existence of the original steady state
of the field and its competition with the oscillatory behavior of $f$
around $f=0$ yields rather unphysical behavior involving large gradients
of the field. For these reasons, this case is not presented in detail here.

\bibitem{eilbeck}  R. K. Dodd, J. C. Eilbeck, J. D. Gibbon, and H. C.
Morris, Solitons and Nonlinear Wave Equations (Academic Press, London, 1982).

\bibitem{weinberg}  B. O'Neill, Semi-Riemannian Geometry with Applications
to Relativity (Academic Press, London, 1983).

\bibitem{ww} In the numerical computations $W$ has been fixed to $W=1$.
By varying $W$, we have found, similarly to the original works of RP,
that its increase slows down the rate at which the presented 
phenomenology occurs.

\bibitem{yannis}  J. Wolff, A.G. Papathanasiou, I.G. Kevrekidis, H.H.
Rotermund, and G. Ertl, Science {\bf 294}, 134 (2001).
\end{references}
\end{document}